\documentclass[runningheads]{llncs}
\usepackage{graphicx}
\usepackage{amsmath,amssymb,amsfonts}
\usepackage{dcolumn}
\usepackage{bm}
\usepackage{xcolor}
\usepackage{qcircuit}
\usepackage{braket}
\usepackage{float}
\usepackage{algorithm}
\usepackage{algorithmicx}
\usepackage{algpseudocode}
\usepackage{mathtools}
\usepackage{hyperref}

\usepackage{amsmath}

\renewcommand{\figureautorefname}{Figure~\negthinspace}

\renewcommand{\sectionautorefname}{Section~\negthinspace}

\begin{document}
\title{Quantum deep Q learning with distributed prioritized experience replay\thanks{The views expressed in this article are those of the authors and do not represent the views of Wells Fargo. This article is for informational purposes only. Nothing contained in this article should be construed as investment advice. Wells Fargo makes no express or implied warranties and expressly disclaims all legal, tax, and accounting implications related to this article.}}
%
%
\author{Samuel Yen-Chi Chen\inst{1}\orcidID{0000-0003-0114-4826}
}
\authorrunning{Chen et al.}
%
\institute{Wells Fargo, New York NY 10017, USA}
\maketitle              
\begin{abstract}
This paper introduces the QDQN-DPER framework to enhance the efficiency of quantum reinforcement learning (QRL) in solving sequential decision tasks. The framework incorporates prioritized experience replay and asynchronous training into the training algorithm to reduce the high sampling complexities. Numerical simulations demonstrate that QDQN-DPER outperforms the baseline distributed quantum Q learning with the same model architecture. The proposed framework holds potential for more complex tasks while maintaining training efficiency.

\end{abstract}
\section{\label{sec:Indroduction}Introduction}

Quantum computing (QC) holds the potential to offer significant computational advantages over classical computing in tasks such as unstructured database search and integer factorization \cite{nielsen2010quantum}. While current noisy intermediate-scale quantum devices \cite{preskill2018quantum} have not yet achieved the highly sought-after quantum supremacy, researchers are actively developing quantum algorithms and applications that can operate effectively on these devices.

Variational quantum algorithms (VQA) \cite{bharti2022noisy} represent a leading approach that shows promise in fields such as optimization and machine learning \cite{mitarai2018quantum}. Among the various VQA-based quantum machine learning algorithms, quantum reinforcement learning (QRL) has been designed specifically for solving sequential decision-making tasks. Several QRL methods have been developed, including quantum deep Q-learning \cite{chen19,lockwood2020reinforcement,skolik2021quantum}, quantum policy gradients \cite{jerbi2021variational,acuto2022variational}, and evolutionary QRL \cite{chen2022variational}. However, training QRL agents can be time-consuming and requires significant computational resources.
This paper presents the \emph{quantum DQN with distributed prioritized experience replay} (QDQN-DPER) framework for training QRL agents. The proposed framework incorporates both asynchronous or distributed training and prioritized experience replay (PER) into quantum deep Q-learning. Furthermore, we introduce a modified loss function to enhance the training process.
We present numerical simulations that demonstrate the superior performance of our proposed QDQN-DPER framework, compared to the asynchronous QDQN without prioritized experience replay or modified loss function, in terms of achieving higher and more stable scores within the considered tasks. Our proposed framework represents a promising avenue for efficient QRL training and future applications.
The paper is organized as follows:
\sectionautorefname{\ref{sec:RelatedWork}} provides a review of recent developments in QRL. In \sectionautorefname{\ref{sec:ReinforcementLearning}}, we briefly introduce the fundamental concepts of RL. \sectionautorefname{\ref{sec:VQC}} presents an overview of variational quantum circuits (VQC), which serves as the foundational building block for QRL models. Our proposed QDQN-DPER framework is introduced in \sectionautorefname{\ref{sec:QDQN_DPER}}. Numerical simulations demonstrating the efficacy of our framework are presented in \sectionautorefname{\ref{sec:Experiments}}, and we conclude our work in \sectionautorefname{\ref{sec:Conclusion}}.
\section{\label{sec:RelatedWork}Related Work}
Quantum reinforcement learning (QRL) \cite{meyer2022survey} can be traced back to \cite{dong2008quantum}, but the framework requires a quantum environment that may not be available in most real-world scenarios. Recent studies have explored the use of Grover-like methods \cite{wiedemann2022quantum,sannia2022hybrid} and quantum linear system solvers for implementing quantum policy iteration \cite{cherrat2022quantum}. In this work, we focus on recent advancements in VQC-based QRL that address classical environments.
In the realm of QRL, two broad categories of approaches can be identified: value-based and policy-based methods. Value-based QRL aims to approximate the optimal value function, while policy-based QRL seeks to directly optimize the policy. The first VQC-based approach in value-based QRL was proposed by Chen et al. \cite{chen19}, which presents a quantum version of deep Q-learning (DQN) algorithm. This method is designed for discrete observation and action spaces, and has been tested on environments such as Frozen-Lake and Cognitive-Radio. Since then, further developments have been made in the area of quantum DQN, including efforts to handle continuous observation spaces, such as Cart-Pole, by Lockwood et al. \cite{lockwood2020reinforcement} and Skolik et al. \cite{skolik2021quantum}. A more recent development in the direction of value-based QRL involves the use of quantum recurrent neural networks (QRNN) as the value function approximator. Chen et al. \cite{chen2022quantum} proposed this approach to address challenges such as partial observability or environments that require a longer memory of previous steps. With the use of QRNN, the value function can be approximated using a quantum neural network that is capable of processing sequential input.
Researchers have developed a range of methods to approximate the value function, including hybrid quantum-classical linear solvers, as discussed in \cite{Chih-ChiehCHEN2020}. In addition, recent work by Heimann et al. \cite{heimann2022quantum} has implemented a modified version of deep Q-networks (DQN), known as Double DQN (DDQN), within the VQC framework to improve agent convergence.
Policy-based QRL methods directly learn policy functions, denoted as $\pi$, in addition to value functions such as the $Q$-function. Such approaches are capable of finding the optimal policy for a given problem. For example, Jerbi et al. \cite{jerbi2021variational} proposed a quantum policy gradient RL framework using the REINFORCE algorithm. Similarly, Hsiao et al. \cite{hsiao2022unentangled} introduced an improved policy gradient algorithm called PPO, demonstrating that even with a small number of parameters, quantum models can outperform their classical counterparts. Jerbi et al. \cite{jerbi2022quantum} showed provable quantum advantages of policy gradient methods. Moreover, Meyer et al. \cite{meyer2022quantum} investigated the impact of post-processing methods for variational quantum circuits (VQC) on the performance of quantum policy gradients. Recently, several improved quantum policy gradient algorithms have been proposed. Schenk et al. \cite{schenk2022hybrid} introduced an actor-critic approach, while Lan et al. \cite{lan2021variational} and Acuto et al. \cite{acuto2022variational} proposed soft actor-critic (SAC) algorithms. Wu et al. \cite{wu2020quantum} introduced a quantum version of deep deterministic policy gradient (DDPG), and Chen et al. \cite{chen2023asynchronous} presented a quantum asynchronous advantage actor-critic (A3C) algorithm. Xiao et al. \cite{xiao2023quantum} introduced a generative adversarial RL in the quantum regime. These modifications aim to further enhance the efficiency and effectiveness of QRL methods.
QRL has been successfully applied in quantum control \cite{sequeira2022variational}. QRL has also been extended to multi-agent settings with promising results \cite{yun2022quantum,yan2022multiagent,yun2022quantum2}. Chen et al. \cite{chen2022variational} were the first to investigate the use of evolutionary optimization for QRL. In their work, multiple agents were initialized and trained in parallel, with the best performing agents being selected as parents to generate the next generation of agents. 

In this work, we build upon previous research in quantum deep Q-learning \cite{chen19,lockwood2020reinforcement,skolik2021quantum} and quantum A3C \cite{chen2023asynchronous} by incorporating asynchronous training and experience replay into quantum Q-learning. While prior approaches to quantum Q-learning have used single-threaded training, our method employs asynchronous training, which offers practical advantages, such as reduced training time using multi-core CPU computing resources and the potential for multiple quantum processing units (QPUs) when available. We also utilize prioritized experience replay (PER), a classical RL technique, to further enhance the training of quantum RL. Our approach is similar to the evolutionary QRL method presented in \cite{chen2022variational}, which also employs parallel computing resources. However, our approach is distinct in that all individual agents share the same global policy and target network. Each agent calculates gradients based on their own experiences and updates the global model asynchronously. There is no need to wait for all agents to finish before calculating fitness and creating the next generation of agents. This feature can further enhance the efficiency of the training process and represent a significant advancement in QRL research.
\section{\label{sec:ReinforcementLearning}Reinforcement Learning}
\emph{Reinforcement learning} (RL) is a machine learning framework in which an \emph{agent} learns to accomplish a given goal by interacting with an \emph{environment} $\mathcal{E}$ in discrete time steps~\cite{sutton2018reinforcement}.
The agent observes a \emph{state} $s_t$ at each time step $t$ and then chooses an \emph{action} $a_t$ from the action space $\mathcal{A}$ based on its current \emph{policy} $\pi$. The policy is a mapping from a specific state $s_t$ to the probabilities of choosing one of the actions in $\mathcal{A}$.
After performing the action $a_t$, the agent gets a scalar \emph{reward} $r_t$ and the state of the following time step $s_{t+1}$ from the environment. The procedure is repeated across a number of time steps until the agent reaches the terminal state or the maximum number of steps permitted. A complete cycle of this is called an \emph{episode}. During the training process, when the RL agent encounters the state $s_t$, it aims to maximize the expected return, which can be mathematically described as the value function at state $s$ under policy $\pi$, $V^\pi(s) = \mathbb{E}\left[R_t|s_t = s\right]$, where $R_t = \sum_{t'=t}^{T} \gamma^{t'-t} r_{t'}$ is the \emph{return}, the total discounted reward from time step $t$. The value function can be further expressed as $V^\pi(s) = \sum_{a\in\mathcal{A}} Q^\pi (s,a) \pi(a|s)$, where the \emph{action-value function} or \emph{Q-value function} $ Q^\pi (s,a) = \mathbb{E}[R_t|s_t = s, a]$ is the expected return of choosing an particular action $a \in \mathcal{A}$ in state $s$ according to the policy $\pi$. The $Q$-learning is RL algorithm to optimize the $Q^\pi (s,a)$ via the following formula
\begin{align}
  Q\left(s_{t}, a_{t}\right) \leftarrow & \, Q\left(s_{t}, a_{t}\right)\nonumber\\
  &+\alpha\left[r_{t}+\gamma \max _{a} Q\left(s_{t+1}, a\right)-Q\left(s_{t}, a_{t}\right)\right].
\end{align}
The $Q^\pi (s,a)$ function can in principle be implemented by a simple table with the size $|\mathcal{S}|\times |\mathcal{A}|$ where $|\mathcal{S}|$ and $|\mathcal{A}|$ denote the size of state space and action space, respectively. However, the states and the actions may be very large discrete spaces or even continuous, making the tabular $Q$ function impractical. Deep neural networks (DNN)  \cite{mnih2015human} and the more recently developed variational quantum circuits (VQC) \cite{chen19} can be used to approximate the $Q$ functions with large state or action spaces.
Directly replacing the tabular $Q$ functions with nonlinear function approximators such neural networks sometimes makes the training of RL agents generally hard to converge \cite{mnih2015human}. Potential reasons are correlated states or observations and optimization towards an unstable target.
\emph{Experience replay} is a successful method for addressing non-i.i.d. sampling. The original setting of experience replay \cite{mnih2015human} is to sample a batch of transitions \emph{uniformly} from the replay memory and optimize the model. Although this approach is viable, it may not be the most efficient method since the model may acquire more knowledge from certain transitions and less from others. For example, experiences that led to large rewards or that resulted in the agent making a significant error are likely to be more informative than experiences that had little impact on the agent's behavior. 
\emph{Prioritized Experience Replay} (PER) \cite{schaul2015prioritized} is a technique for experience replay which involves replaying transitions with a high expected learning progress, as determined by the size of their temporal-difference (TD) error, more frequently than others.
The probability of sampling the transition $i$ is defined to be $P(i)=\frac{p_i^\alpha}{\sum_k p_k^\alpha}$ where $p_i>0$ is the priority of transition $i$. The parameter $\alpha$ controls how much the prioritization is used. It means sampling transitions uniformly from the replay memory when $\alpha=0$. 
To estimate expected values with stochastic updates, the updates should match the distribution of the underlying expectation. Prioritized replay can introduce bias and alter the distribution, affecting the convergence of estimates. To fix this, importance-sampling weights can be used to reduce bias.
The importance sampling (IS) weights are defined as $w_i=\left(\frac{1}{N} \cdot \frac{1}{P(i)}\right)^\beta$ where $N$ is the size of replay memory. 
%
For stability reasons, we always normalize weights by $1 / \max _i w_i$.
The \emph{Distributed Prioritized Experience Replay} (DPER) algorithm, introduced in \cite{horgan2018distributed}, extends the PER algorithm to a distributed setting, enabling multiple agents to interact with the environment in parallel. In the original formulation of DPER, multiple actors generated experience concurrently, and a shared replay memory was used. The learner collected the experience from the actors and trained the model using experience sampled from the shared memory. In this paper, we propose a modified version of DPER that eliminates the need for uploading local experiences. Instead, each local agent maintains its own replay memory and uses locally sampled experiences to update the global shared model.
\section{\label{sec:VQC}Variational Quantum Circuit}
\begin{figure}[htbp]
\includegraphics[width=1\linewidth]{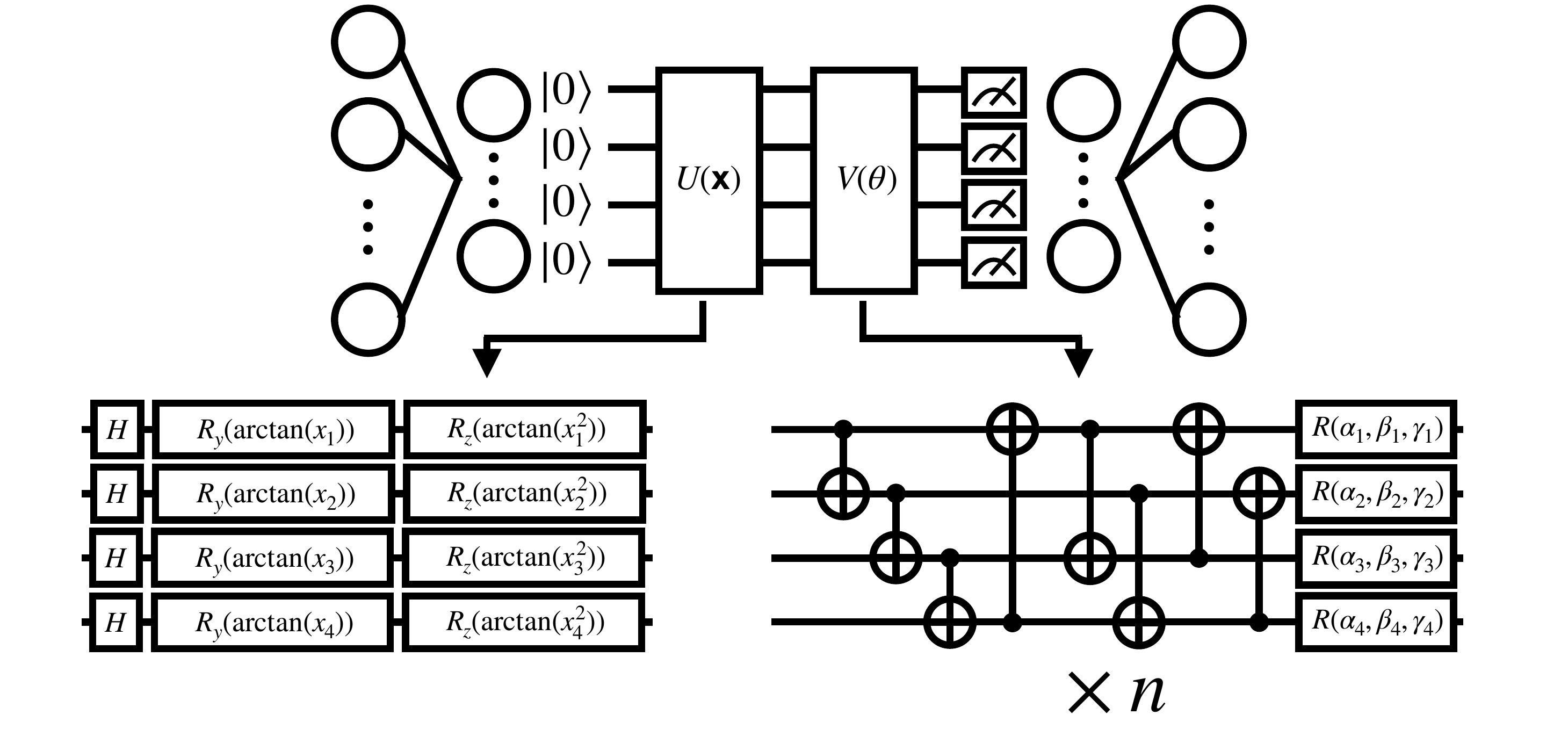}
\caption{{\bfseries Hybrid variational quantum circuit (VQC) architecture.} (Upper part) The hybrid VQC architecture includes a VQC and classical neural networks (NN) before and after the VQC. NN can be used to reduce the dimensionality of the input data and refine the outputs from the VQC. (Lower part) The encoding circuit $U(\mathbf{x})$ includes Hadamard gate and single qubit rotations $R_{y}$ and $R_{z}$. The variational circuit includes CNOT gates to entangle qubits and parameterized unitary rotation gates $R(\alpha, \beta, \gamma)$. The variational circuit block can be repeated $n$ times to increase the number of quantum parameters.}
\label{fig:hybrid_vqc}
\end{figure}

Variational quantum circuits (VQCs), also known as parameterized quantum circuits (PQCs), are quantum circuits that allow for the adjustment of their parameters. These parameters can be optimized using classical machine learning methods, including gradient-based and non-gradient-based techniques. \figureautorefname{\ref{fig:hybrid_vqc}} shows a typical illustration of a VQC, which consists of three primary components: an encoding circuit, a variational circuit, and a quantum measurement layer. The encoding circuit, denoted as $U(\textbf{x})$, transforms classical values into a quantum state, while the variational circuit, denoted as $V(\theta)$, is the learnable portion of the VQC. The quantum measurement layer extracts information from the circuit by repeatedly executing the circuit to obtain the expectation values of each qubit. The most commonly used expectation value is the Pauli-$Z$ expectation value, which is obtained as a float value instead of a binary integer. To process the values obtained from the circuit, additional VQCs or classical components such as deep neural networks or tensor networks can be employed. These enhanced VQCs are referred to as \emph{dressed} VQCs.

End-to-end training of the whole VQC-based model is possible through gradient-based \cite{chen2021end,qi2021qtn} or gradient-free methods \cite{chen2022variational}. Gradient-based methods involve representing the entire model as a directed acyclic graph (DAG), which enables back-propagation to be applied. The success of end-to-end optimization depends on the ability to calculate quantum gradients, such as the \emph{parameter-shift} rule \cite{mitarai2018quantum}. VQC-based quantum machine learning (QML) models have demonstrated success in several areas, including classification \cite{mitarai2018quantum,qi2021qtn,chehimi2022quantum,chen2021federated,chen2021end}, natural language processing \cite{yang2020decentralizing,yang2022bert,di2022dawn}, and sequence modeling \cite{chen2020quantum,chen2022reservoir}.

\section{\label{sec:QDQN_DPER}Quantum DQN-DPER}
The proposed QDQN-DPER includes a global policy network, global target network and an array of parallel agents. 
The global policy and target networks are dressed VQCs following the design we described in \sectionautorefname{\ref{sec:VQC}}. The observations agents get from the environment will be first processed by the classical NN before entering the VQC. Then the output from the VQC will be further processed by another VQC to generate the final outputs for $Q$ functions. 
Each of the agent holds their own prioritized experience replay memory $\mathcal{D}$ with a fixed length $|\mathcal{D}|$. The replay memories save the trajectories $(s[t:t+k-1], a[t:t+k-1], r[t:t+k-1], s[t+1:t+k])$ encountered by the agent in which there are up to $t = S$ steps. Here $s[t:t+k-1]$ denotes the list $s_{t} \cdots s_{t+k-1}$ The action choice follows the $\epsilon$-greedy strategy. There is a global counter $\mathcal{C}$ to keep track of the total number of steps passed. The global target network will be updated from the global policy network every $C$ steps.
The target of values at time $t$ is $\tilde{v}(t) = r_t+\gamma \max _a Q\left(s_{t+1}, a, \theta^{-}\right)$ and the prediction from the model is $Q\left(s_{t}, a_{t}, \theta\right)$, where $\theta^{-}$ and $\theta$ represent the model parameter of the target net and policy net, respectively. The original definition of loss $L$ is $L = \left[ r_t+\gamma \max _a Q\left(s_{t+1}, a, \theta^{-}\right) - Q\left(s_{t}, a_{t}, \theta\right)\right]^2$. Here we propose a modified version of loss in the following way: consider the $n$-step trajectory sampled from the memory. For time $t_{k} \cdots t_{k + (n - 1)} $, we can calculate the target of values as $\tilde{v}(t_{k+i}) = r_{t_{k+i}}+\gamma \max _a Q\left(s_{t_{k+i+1}}, a, \theta^{-}\right)$ and the predictions from the model as $Q\left(s_{t_{k+i}}, a_{t_{k+i}}, \theta\right)$. We define the loss matrix as,
\begin{equation}
    \begin{bmatrix}
        \tilde{v}(t_{k})- Q\left(s_{t_{k}}, a_{t_{k}}, \theta\right)& \cdots & \tilde{v}(t_{k})- Q\left(s_{t_{k+n-1}}, a_{t_{k+n-1}}, \theta\right) \\
        \vdots & \ddots & \\
        \tilde{v}(t_{k+n-1})- Q\left(s_{t_{k}}, a_{t_{k}}, \theta\right)& \cdots & \tilde{v}(t_{k+n-1})- Q\left(s_{t_{k+n-1}}, a_{t_{k+n-1}}, \theta\right) \\
        
\end{bmatrix}
\end{equation}
The diagonal elements are the original TD terms: $r_{t_{k+i}}+\gamma \max _a Q\left(s_{t_{k+i+1}}, a, \theta^{-}\right) - Q\left(s_{t_{k+i}}, a_{t_{k+i}}, \theta\right)$.
The mean square of the matrix $\tilde{L} = \frac{1}{n^2}\sum_{p = 0}^{n-1}\sum_{q = 0}^{n-1} (\tilde{v}(t_{k + p}) - Q(s_{t_{k + q}}, a_{t_{k + q}}, \theta))^{2}$ is used as the scalar loss for calculating the gradients. This loss is also used to evaluate the priorities of the trajectories in the replay memory.
We summarize the proposed QDQN-DPER in Algorithm.\ref{QDQN_DPER_alg}.

\section{\label{sec:Experiments}Experiments}
\subsection{Environment}
\begin{figure}[htbp]
\center
\includegraphics[width=0.5\linewidth]{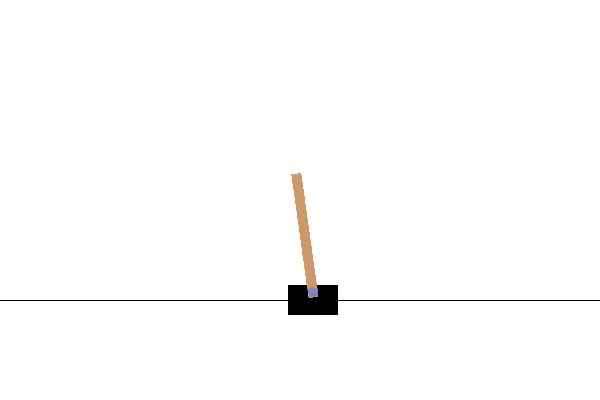}
\caption[Environment: Cart-Pole]{{\bfseries The Cart-Pole environment from OpenAI Gym.}
 }
\label{CartPole_Env}
\end{figure}
Cart-Pole is a standard evaluation environment in OpenAI Gym for RL models \cite{brockman2016openai} (see \figureautorefname{\ref{CartPole_Env}}). The original version has a frictionless track with a fixed joint connecting a pole to a cart. The goal is to maintain the pendulum near its starting position by moving the cart left or right. The RL agent receives four-dimensional observations $s_t$ of cart position, velocity, pole angle, and pole tip velocity at each time step. The agent is rewarded with +1 if the pole is nearly upright and episodes end if the pole tilts more than 15 degrees from vertical or if the cart moves more than 2.4 units away from the center.
In this study, we investigate a modified version of the Cart-Pole environment, known as CartPoleMod \cite{AadityaRavindran}, which takes into account the effects of friction on the system. We have determined the frictional coefficients for the cart and pole to be $5 \times 10^{-4}$ and $2 \times 10^{-6}$, respectively. Moreover, we have examined several noise configurations, ranging from v0 to v3, within the CartPoleMod environment. Specifically, v0 represents an environment without any noise, whereas v1 and v2 introduce $5\%$ and $10\%$ uniform actuator noise, respectively. In addition, v3 incorporates $5\%$ uniform sensor noise.
\subsection{Hyperparameters and Model Size}
In the proposed Quantum DQN-DPER, we use the RMSprop optimizer with learning rate $1 \times 10^{-3}$, $\alpha = 0.99$ and $\epsilon = 1 \times 10^{-8}$. 
The discount factor $\gamma$ is set to be $0.9$. Within each local agent, trajectories are saved to the process-specific replay memory every $S = 5$ steps and then a batch $B = 4$ of trajectories are sampled to calculate the loss and gradients. The parameters $\alpha, \beta$ controlling the prioritized sampling are set to be $0.6$ and $0.4$, respectively. The target network are updated every $C = 2000$ global steps. The process-specific $\epsilon$ for $\epsilon$-greedy starts from $1$ with a decay rate $0.9999$ and minimum value $0.001$. We perform simulation for each environment and setting with 50,000 episodes.
In this study, the VQC has been configured with a specific number of qubits, namely 8, and two variational layers have been incorporated into its design. As a result, the VQC employs a total of 48 quantum parameters, derived from the product of the number of qubits (8), the number of circuit layers (2), and the number of adjustable parameters per unitary rotation gate (3). It is noteworthy that the VQC architecture remains consistent across all testing environments analyzed in this research. In the context of classical benchmarks, we examine models that bear a strong resemblance to the dressed VQC model. To be specific, we maintain the classical model's architecture, as depicted in \figureautorefname{\ref{fig:hybrid_vqc}}, while substituting the 8-qubit VQC with a single layer having input and output dimensions of 8. This results in an architecture that closely resembles the quantum model, with a parameter count that is also highly comparable. We consider the classical asynchronous Q learning with and without PER.
As delineated in \sectionautorefname{\ref{sec:QDQN_DPER}}, this study utilizes single layer neural networks for preprocessing data prior to and subsequent to the VQC. The input dimensions of the networks preceding the VQC are determined by the specific environments that the agent is intended to solve. In the context of this research, we examine the CartPoleMod environment, which is characterized by a four-dimensional observation vector.
\subsection{Results}
\begin{figure}[htbp]
\includegraphics[width=1\linewidth]{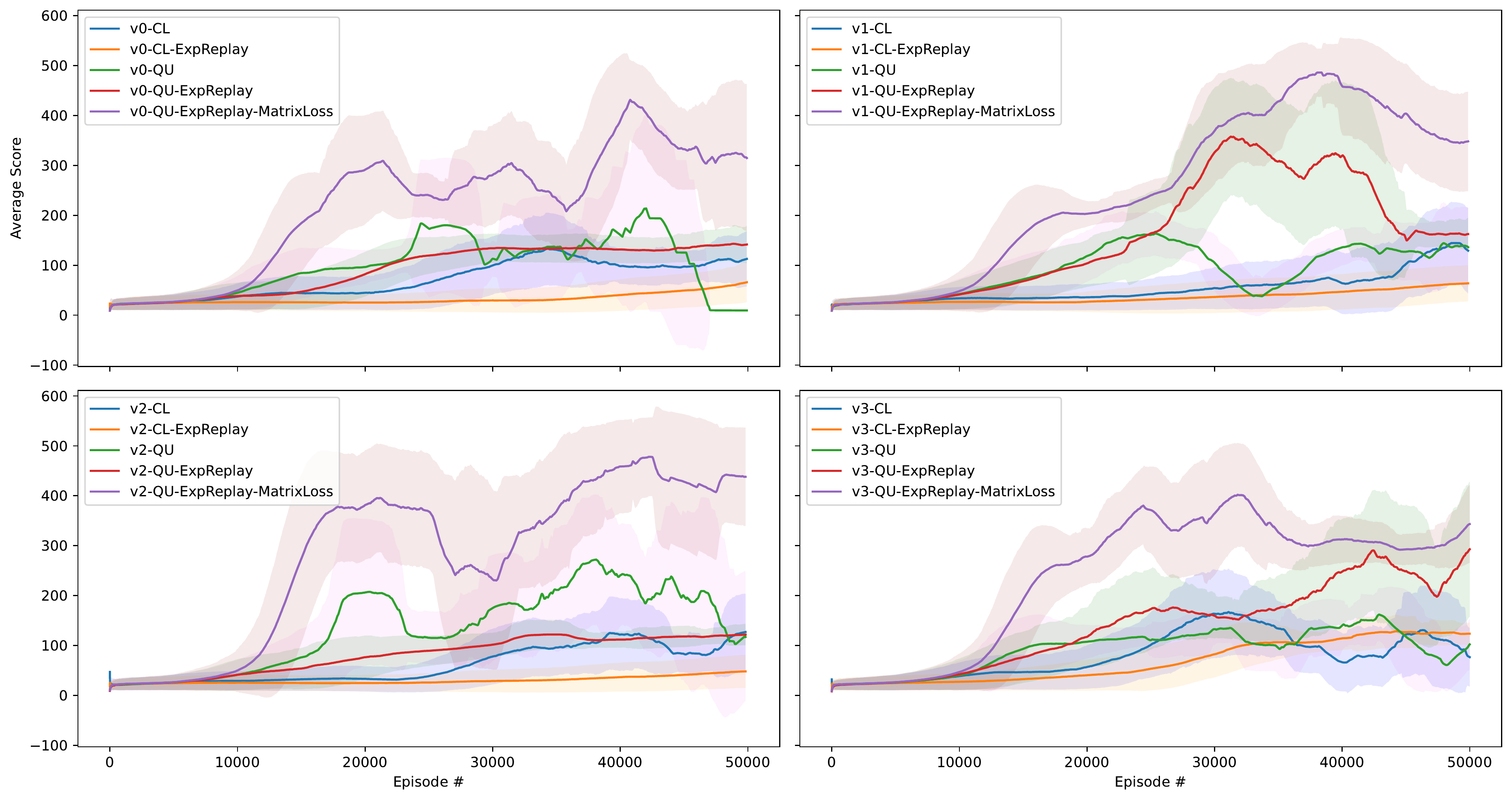}
\caption{{\bfseries Results: Quantum Asynchronous Q learning in the CartPoleMod environment.}  }
\label{fig:results_cartpolemod}
\end{figure}
The simulation results of all experimental configurations are shown in the \figureautorefname{\ref{fig:results_cartpolemod}}. We calculate the mean and standard deviation of scores in the past 5,000 episodes to demonstrate the trends. Our study aimed to evaluate the effectiveness of various training methods for asynchronous quantum Q learning models on different CartPoleMod environments. In the first testing case, CartPoleMod-v0, we observed that the model equipped with both prioritized experience replay and matrix loss consistently outperformed the other two quantum models (with experience replay and without experience replay). Notably, we found that the matrix loss was crucial in achieving convergence within 50,000 episodes. Furthermore, in the quantum models, the prioritized experience replay further improved the model's performance, resulting in higher and more stable scores than the model without experience replay.
In the second testing case, CartPoleMod-v1, we observed a similar pattern to v0. We noted that the model with both matrix loss and prioritized experience replay outperformed the model without the aid of matrix loss and the model without both. Our results demonstrated that the combination of prioritized experience replay and matrix loss played a crucial role in stabilizing the QRL training.
In cases v2 and v3, we observed the same pattern as in v0, with the model equipped with both prioritized experience replay and matrix loss consistently outperforming the other models. 
Among all testing cases, quantum models with prioritized experience replay beat the classical models with similar model size (number of parameters) and architecture.
Overall, our results were consistent across all CartPoleMod environments considered, leading us to conclude that the combination of prioritized experience replay and matrix loss is an effective approach to significantly enhance the training of asynchronous quantum Q learning models.
\section{\label{sec:Conclusion}Conclusion}
In this study, we investigate the efficacy of an asynchronous training framework for quantum Q-learning agents with distributed prioritized experience replay. Through numerical simulations, we demonstrate that the proposed approach yields superior performance compared to traditional quantum deep Q-network models that lack the prioritized replay memory or modified loss function in the benchmark tasks considered. The results suggest that the proposed approach enables efficient training of quantum RL agents by leveraging parallel computing resources and has the potential for various real-world applications.
\begin{center}
\scalebox{0.7}{
\begin{minipage}{\linewidth}
\begin{algorithm}[H]
\begin{algorithmic}
\State \textbf{Define} the trajectory length parameter $S$
\State \textbf{Define} the target network update parameter $C$
\State \textbf{Assume} global shared hybrid VQC policy network parameter $\theta$
\State \textbf{Assume} global shared hybrid VQC target network  parameter $\theta^{-}$
\State \textbf{Assume} global shared episode counter $T = 0$
\State \textbf{Assume} global shared steps counter $Y = 0$

\State \textbf{Initialize} the process-specific replay memory $\mathcal{D}$ to capacity $|\mathcal{D}|$
\State \textbf{Initialize} the process-specific steps counter $Z = 0$

\While{$T < T_{max}$}
    \State Initialize the trajectory record buffer $\mathcal{M}$
    \State Initialise state $s_1$ and encode into the quantum state
    \While{Not done}
    	\State With probability $\epsilon$ select a random action $a_t$
    	\State otherwise select $a_t = \max_{a} Q^*(s_t, a; \theta)$ from the output of the (global) policy network
    	\State Execute action $a_t$ in emulator and observe reward $r_t$ and next state $s_{t+1}$
    	\State Store transition $\left(s_t,a_t,r_t,s_{t+1}\right)$ in $\mathcal{M}$
            \If {$Z \mod S = 0$ or done}
                \State Store episode record $\mathcal{M}$ to $\mathcal{D}$ and re-initialize $\mathcal{M}$.
        	\State Sample random batch of trajectories $\mathcal{T}$ and the corresponding $w$ from $\mathcal{D}$
                \State Calculate the losses $l_{1} \cdots l_{B}$ for each sampled trajectory.
                \State Update the probability of sampled trajectories in $\mathcal{D}$ with newly calculated losses.
                \State Calculate the weighted average loss $L = \sum_{i} w_{i} l_{i}$
                \State Perform a gradient descent step on $L$
            \EndIf
            \If {$Y \mod C = 0$}
                \State Update the target network $\theta^{-}$ with $\theta$.
            \EndIf
            \State Update $\epsilon$
    \EndWhile
    
\EndWhile

\end{algorithmic}
\caption{Quantum DQN with Distributed Prioritized Experience Replay (QDQN-DPER) (algorithm for each learner process)}
\label{QDQN_DPER_alg}
\end{algorithm}
\end{minipage}
}
\end{center}

\bibliographystyle{splncs04}
\bibliography{bib/qrl,bib/classical_rl,bib/qml_examples,bib/tool,bib/qc,bib/vqc}

\end{document}